\documentclass[pre,aps,showpacs,byrevtex,amsmath,amssymb,onecolumn]{revtex4}
\usepackage{graphicx}

\begin{document}

\title{Nucleosome interactions in chromatin: fiber stiffening
and hairpin formation}

\author{Boris Mergell$^{\text{1}}$}
\email{mergell@mpip-mainz.mpg.de}
\author{Ralf Everaers$^{\text{2}}$}
\email{everaers@mpipks-dresden.mpg.de}
\author{Helmut Schiessel$^{\text{1}}$}
\email{heli@mpip-mainz.mpg.de}

\affiliation{$^{\text{1}}$Max-Planck-Institut f\"ur
Polymerforschung,
 Postfach 3148, D-55021 Mainz,
 Germany}
\affiliation{$^{\text{2}}$Max-Planck-Institut f\"ur die Physik komplexer Systeme,
 N\"othnitzer Stra{\ss}e 38, D-01187 Dresden,
 Germany}

\date{\today}

\begin{abstract}

We use Monte Carlo simulations to study attractive and excluded
volume interactions between nucleosome core particles in 30
nm-chromatin fibers. The nucleosomes are treated as disk-like
objects having an excluded volume and short range attraction
modelled by a variant of the Gay-Berne potential. The nucleosomes
are connected via bendable and twistable linker DNA in the crossed
linker fashion. We investigate the influence of the nucleosomal
excluded volume on the stiffness of the fiber. For parameter
values that correspond to chicken erythrocyte chromatin we find
that the persistence length is governed to a large extent by that
excluded volume whereas the soft linker backbone elasticity plays
only a minor role. We further find that internucleosomal
attraction can induce the formation of hairpin configurations.
Tension-induced opening of such configurations into straight
fibers manifests itself in a quasi-plateau in the force-extension
curve that resembles results from recent micromanipulation
experiments. Such hairpins may play a role in the formation of
higher order structures in chromosomes like chromonema fibers.

\end{abstract}

\pacs{87.14.Gg,87.15.Aa,87.15.La,61.41.+e}

\maketitle


\section{Introduction}

DNA of all eucaryotic organisms is wrapped around millions of
cylindrical protein spools, so-called histone octamers. Each
complex has a radius of 5 nm and a height of 6 nm and -- together
with the stretch of linker DNA connecting to the next such spool
-- is called nucleosome, the basic unit of the chromatin complex
\cite{Holde_89}. In a next step of DNA compaction the string of
nucleosomes organizes itself into the chromatin fiber. For low
salt concentrations a 'beads-on-a-string' structure is observed,
sometimes referred to as the 10-nm fiber \cite{Thoma_jcellb_79}.
For higher salt concentrations ($>$ 40 mM) the fiber appears to
thicken into a condensed structure with a diameter of roughly $30$
nm \cite{Widom_jmb_86}. The degree of compaction also depends
strongly on the presence of linker histones. They cause the in-
and outcoming DNA to form a short stem-like structure
\cite{Bednar_pnas_98}. In the absence of linker histones the
entry-exit angle of the in- and outcoming DNA is larger, leading
to more open structures.

While the structure of the nucleosome core particle (the protein
spool with the 2 turns of wrapped DNA) is known up to atomistic
resolution \cite{Luger_nat_97} there is still considerable
controversy about the details of the structure of the 30-nm
chromatin fiber
\cite{Holde_89,Widom_arbc_89,Holde_jbc_95,Holde_pnas_96,Langowski_03}.
There are essentially two competing classes of models: (i) the
solenoid models \cite{Thoma_jcellb_79,Finch_pnas_76,Widom_cell_85}
and (ii) the zig-zag- or crossed-linker models
\cite{Woodcock_pnas_93,Bednar_pnas_98,Schiessel_bpj_01}. In the
solenoid models one assumes that the successive nucleosomes form a
helix with the normal vectors of the protein spools (the axis of
the superhelical wrapping path of the DNA) being perpendicular to
the solenoidal axis. The entry-exit sides of the nucleosomal DNA
face inward towards the solenoidal axis and the linker DNA must
bend in order to connect neighboring nucleosomes. In contrast the
linker DNA in crossed-linker models remains straight and connects
nucleosomes that sit on opposite sides of the fiber.

Chromatin fibers have been studied for various salt concentrations
using electron cryo-microscopy
\cite{Bednar_jcb_95,Bednar_pnas_98}, atomic force microscopy
\cite{Leuba_pnas_94,zlatanova_bpj_98}, neutron scattering and
scanning transmission electron microscopy \cite{Gerchman_pnas_87}.
Structural parameters such as the mass density (number of
nucleosomes per 11 nm) and the linker entry-exit angle are
measured to characterize the state of compaction.  For low salt
concentrations all these studies support the picture of an open
zig-zag-like fiber structure. Similarly the force-extension curves
of single chromatin fibers measured under these conditions
\cite{Cui_pnas_00} (for a recent review see
Ref.~\cite{Zlatanova_jmb_03}) are in good agreement with the
results from computer simulations \cite{Katritch_jmb_00} and
analytical approaches
\cite{Schiessel_bpj_01,Haim_pre_01,Haim_physicaA_02} based on a
crossed-linker geometries of the fiber (for a recent review see
Ref.~\cite{Helmut_03}).

The situation is less clear at physiological salt concentrations.
The internal structure of the dense 30-nm fiber  could not yet be
resolved despite enormous experimental efforts, especially X-ray
diffraction studies (cf. Ref. \cite{zlatanova_jmb_95} for a
critical discussion).  The interpretation of related studies on
di- and trinucleosomes is also still controversial
\cite{Holde_pnas_96,Bednar_jcb_95,Yao_pnas_90,Butler_jmb_98}. Even
less is known about how chromatin folds into chromosomes on larger
scales. Theories for the elastic properties of the DNA linker
backbone predict rather flexible fibers with persistence lengths
on the order of 10-20 nm (compared to 50 nm for uncomplexed DNA).
In contrast, Langowski and coworkers
\cite{Munkel_pre_98,Wedemann_bpj_02} have presented convincing
evidence from experiments and simulations for rather large
persistence length of the order of 300 nm. At the same time, the
quasi-plateau at 5 pN that Cui and Bustamante observed in their
stretching experiments at physiological salt concentrations points
to rather delicate features in the fiber structure (for
comparison, the B-S transition in DNA is observed at a critical
force of 65 pN \cite{Smith_cosb_00}).

In the present paper we use computer simulations to study the
consequence of internucleosome attractive and excluded volume
interactions in dense crossed-linker fibers. Our chromatin fiber
model is depicted on the lhs of Fig.~\ref{fig:modelchrom}. We
model the protein spools as ellipsoidal disks and assume that
linker histones induce the in- and outcoming DNA to form stem-like
structures \cite{Bednar_pnas_98}. Our opening and rotation angles
$\pi-\theta$ and $\phi$ are compatible with experimentally
observed values for the 30 nm fibers
\cite{Bednar_pnas_98,Widom_pnas_92} (see the fiber ``10'' on the
rhs of Fig.~\ref{fig:modelchrom}). As a consequence, the {\em
local} fiber geometry remains basically unchanged when we vary the
strength of the attractive interaction. Local
condensation-decondensation transitions in chromatin
fibers~\cite{Schiessel_bpj_01,Helmut_03} will be investigated in a
forthcoming study \cite{everaers_prep}.

With respect to the degree of coarse-graining of the nucleosome
structure our model is comparable to the one employed by Katritch
et al.~\cite{Katritch_jmb_00}. The main difference is their use of
quenched, randomly distributed $\phi$-angles along the chain.
Since we are interested in rather generic features of the dense 30
nm fiber with a narrow distribution of linker lengths (and hence
twist angles) \cite{Widom_pnas_92}, we focus on the ideal case of
a constant $\phi$-angle for all linkers. Furthermore we follow
Wedemann and Langowski \cite{Wedemann_bpj_02} in modelling the
nucleosomes as ellipsoids (as opposed to spheres as in
Ref.\cite{Katritch_jmb_00}) interacting via a Gay-Berne potential.
However, we have not included Debye-H\"uckel interactions between
different parts of the linker DNA~\cite{Wedemann_bpj_02}, since
the Debye-H\"uckel screening length is smaller than the diameter
of the DNA double helix at physiological salt concentrations. As a
consequence, intra and inter-linker electrostatic interactions can
be accounted for by renormalized bending
rigidities~\cite{Barrat_acp_96} and opening angles $\pi-\theta$
\cite{Schiessel_el_02} respectively.

Problematic for all attempts to model chromatin are the soft parts
of the nucleosome. The charges on the linker histones H1/H5 and
the histone tails are under biochemical control, allowing the cell
to regulate the stem formation and the attractive interactions
between nucleosomes \cite{Holde_pnas_96,Mangenot_bpj_02}. While
simulations investigating this "tail-bridging" effect
\cite{Mangenot_epje_02,Podgornik_jcp_03} between nucleosomes are
on the way \cite{muhlbacher_03}, we prefer a generic nucleosome
interaction potential to a model which neglects the tails but
accounts in detail  for the surface charges on the nucleosome core
\cite{Beard_structure_01}. Similarly, little is known about the
stem structure (see the cryo micrographs in Ref.
\cite{Bednar_pnas_98}) and for simplicity we have assumed that the
short axis of the nucleosomal disk (the axis of the superhelical
wrapping path) is oriented perpendicular to the plane defined by
the in- and outgoing DNA linker.

The paper is organized as follows. Section \ref{sec:methods}
introduces our chromatin model and the methods used. We present
our results for the mechanical fiber properties in section
\ref{sec:results}. Section \ref{sec:discussion} provides a
discussions of the observed hairpin structures and possible
biological implications of our findings. Finally we give a
conclusion in Section \ref{sec:conclusion}.

\section{Model and Methods}
\label{sec:methods}

\subsection{Definition of the chromatin model}

\begin{figure}[t]
  \begin{center} {\includegraphics[angle=0,width=0.8\linewidth]{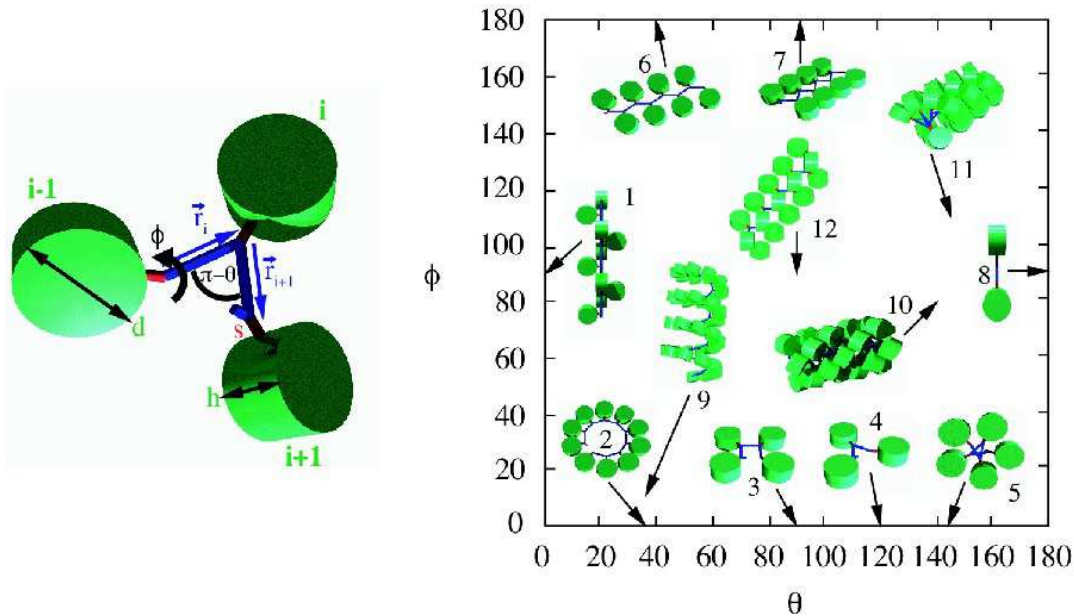}}
  \end{center}
  \caption{Lhs: Portion of the model fiber including three nucleosomes
  (represented in this figure as green {\em cylinders} instead of ellipsoids to
  facilitate the identification of the nucleosome orientation) connected via
  stems (red) to the DNA linkers (blue). Also indicated are the two underlying
  angles: the deflection angle $\theta$ and the rotational angle $\phi$. Rhs:
  Examples of two-angle fibers with the arrows denoting their
  position in the $(\theta,\phi)$-plane. } \label{fig:modelchrom}
\end{figure}

Our chromatin fiber model is depicted on the lhs of
Fig.~\ref{fig:modelchrom}. We model the protein spools as
ellipsoidal disks with a diameter of 10 nm and a height of 6 nm
corresponding to the experimental values
\cite{Livolant_bpj_97,Luger_nat_97}. We assume that the in- and
outcoming DNA are glued together in rigid stem-like structures
that mimic the nucleosomal structures observed in the electron
cryo micrographs in the presence of linker histones
\cite{Bednar_pnas_98}. The stem portion is assumed to end 7 nm
from the center of the corresponding protein spool. From this
point on the entering and exiting linker DNA portions of length
$B$ connect to the next nucleosomes.

The presence of a stem has to be taken into account when one
calculates the length of the free linker DNA from the nucleosomal
DNA repeat length, a number that varies not only from organism to
organism but even from tissue to tissue of the same organism
\cite{Holde_89}. It is known that 147 base pairs (bp) are closely
associated with the protein spool wrapping the histone core in
1.65 turns \cite{Luger_nat_97}. We assume here that the presence
of the linker histone forces the DNA to wrap two full turns
corresponding to 177 bp. Together with the 2 nm stem this makes
189 bp that are associated with the core and linker histones. For
instance, in the case of chicken erythrocyte chromatin the
nucleosomal repeat length amounts to about 210 bp \cite{Holde_89}
so that there are roughly 21 bp, i.e. $B=7.14$ nm, of free linker
DNA. In the following simulation runs we will always use this
value for $B$.

There are two angles that determine the fiber geometry: the
deflection angle $\theta$ and the rotational angle $\phi$. The
former angle characterizes the entry/exit angle $\pi - \theta$ of
the linker DNA at the stem, the latter angle describes the
rotational setting of the nucleosomal disks with respect to the
DNA. Some example configurations are depicted on the rhs of
Fig.~\ref{fig:modelchrom} with the arrows pointing to their
location in the $(\theta,\phi)$-space. Note that some
configurations are forbidden since they would lead to overlapping
nucleosomes; a systematic investigation of the boundary between
allowed and forbidden structures will be provided in a forthcoming
study \cite{everaers_prep}.

In the present study we use a canonical value of $\theta =
145^{\circ}$ which has been estimated experimentally
\cite{Bednar_pnas_98} at 80 mM salt concentration.
Having fixed $B$ and $\theta$ we have chosen values of $\phi$ that lead to a
reasonable nucleosome line density.  Based on Eq.~(78) in
Ref.~\cite{Helmut_03} we expect 6.5 nucleosomes per 11 nm for
$\phi=100^{\circ}$ and 6.1 nucleosomes per 11 nm for $\phi=110^{\circ}$
compared to experimental estimates of 6 - 7 nucleosomes per 11 nm
\cite{Gerchman_pnas_87,Bednar_pnas_98}.

The only non-rigid elements in our model are the linker DNA
portions. We discretize the linkers into four segments in order to
allow for the bending and torsional deformations. To each segment
$i$ we attach a local set of basis vectors $\{{\mathbf
t}_i,{\mathbf n}_i,{\mathbf b}_i\}$ where ${\mathbf t}_i$ denotes
the tangent vector, ${\mathbf n}_i$ the normal, and ${\mathbf
b}_i$ the binormal vector. The elastic energy of the linkers is
then described by
\begin{equation}
  \frac{{\cal H}_{el}}{k_BT} = \frac{\widetilde{l}_p}{2b}\sum_{i=1}^{4(N-1)+2}
  (\beta_i-\beta_{sp})^2 + \frac{\widetilde{l}_{Tw}}{2b}\sum_{i=1}^{4(N-1)+2}
  (\tau_i-\tau_{sp})^2,
\end{equation}
where $\widetilde{l}_p$ and $\widetilde{l}_{Tw}$ are the DNA bending and twist
persistence lengths respectively, $N$ is the number of nucleosomes in our
fiber and $b$ denotes the segment length (i.e. $B=4b$).
$\beta_i=\arccos({\mathbf t}_i\cdot{\mathbf t}_{i+1})$ refers to the bending
angle between two neighboring segments and $\tau_i$ denotes the twist
angle. The spontaneous bending angle $\beta_{sp}$ is equal to $\theta$ for
those segment pairs which are connected to a stem and is zero otherwise.
$\tau_{sp}=\phi/3$ everywhere (except at the kink where $\tau_{sp}=0$)
enforces the right-handed helicity of the DNA which in turn gives rise for the
fiber twist angle $\phi$.

The position and orientation of the nucleosomes is calculated from
the linker positions with the help of a set of three orthonormal
basis vectors $\{{\mathbf T}_i,{\mathbf N}_i,{\mathbf B}_i\}$.
${\mathbf N}_i$ is the normal vector perpendicular to the disk
plane that we assume to be given by
\begin{equation}
  {\mathbf N}_i = \frac{{\vec r}_i \times {\vec
      r}_{i+1}}{|{\vec r}_i \times {\vec r}_{i+1}|}.
\end{equation}
Here the set $\{{\vec r}_i\}$ denotes the vectors that connect the
stems of neighboring nucleosomes (see Fig.~\ref{fig:modelchrom}).
${\mathbf B}_i$ points from the tip of the stem towards the center
of the nucleosomal disk, i.e.
\begin{equation}
  {\mathbf B}_i = \frac{{\vec r}_i - {\vec
      r}_{i+1}}{|{\vec r}_i - {\vec r}_{i+1}|}.
\end{equation}
Finally, ${\mathbf T}_i$ is defined as
\begin{equation}
  {\mathbf T}_i = {\mathbf N}_i \times {\mathbf B}_i.
\end{equation}

For the interactions of the nucleosome core particles
we use the same variant of the Gay-Berne (GB) potential
for ellipsoids of arbitrary shape \cite{Berardi_cpl_98,ralf_GB}
as in a recent study of a stacked-ellipsoid model of
DNA \cite{mergell_pre_03}. Here we set the
structure matrix (Eq. (15) in Ref. \cite{mergell_pre_03})
equal to

\begin{equation}
{\cal{S}} =
\begin{pmatrix}
  h/2 & 0 & 0 \\
  0 & d/2 & 0 \\
  0 & 0 & d/2
\end{pmatrix},
\end{equation}
where $d=1.67h$ is the diameter of the nucleosome core particle
with a canonical value for the core particle height of $h=6$ nm
\cite{Luger_nat_97}. The effective distance of closest approach
$h_{12}$ between two ellipsoids is calculated using Eqs. (17)-(19)
in Ref. \cite{mergell_pre_03}. In cases where we study purely
repulsive hard core interactions, we reject all Monte Carlo moves
leading to values $h_{12}<0$. In cases where we study attractive
interactions, the distance dependent part of the interaction
potential Eq. (14) in Ref. \cite{mergell_pre_03} is given by a
modified Lennard-Jones potential

\begin{equation}\label{eq:U0}
U_{\mathrm r} = 4\epsilon_{\mathrm GB}
\left( \left(\frac h {h_{12}+h}\right)^{12} -
       \left(\frac h {h_{12}+h}\right)^{6}
\strut\right)
\end{equation}
The orientation dependent parts are defined in Eqs. (20)-(25) in
Ref. \cite{mergell_pre_03}.  This parameterization of the GB potential leads to
a preferred lateral spacing of about $7.0\mbox{ nm}$ and a vertical spacing of
about $11.0\mbox{ nm}$ of the disks. The values are in good agreement with
experimental results for columnar phases of nucleosome core particles
\cite{Livolant_bpj_97}.  In this paper we characterize the strength of the
nucleosome-nucleosome attraction by the depth of the attractive well for two
parallel disks $\epsilon = 1.41 \epsilon_{GB}$ instead of the prefactor
$\epsilon_{GB}$ in Eq.~(\ref{eq:U0}).

\subsection{Methods}
\label{sec:mc}

We use a Monte-Carlo scheme to simulate the chromatin fiber which
relies on the following three moves: (i) a local move where one
randomly chooses  a nucleosome that is rotated around an axis
determined by two points on the in- and out-coming linker DNA by a
small random angle, (ii) a non-local pivot move where a random
segment point is chosen at which the shorter part of the chain is
rotated around a random axis by a random angle and (iii) a
non-local crankshaft move where two random points along the DNA
segments define the axis of rotation around which the inner part
of the chain is rotated. The moves are accepted or rejected
according to the Metropolis scheme \cite{Metropolis_jcp_53}.

We simulated either fibers with $N=50$ (simulations with applied
stretching force) or $N=100$ nucleosomes corresponding to 10 and
20 kbp of DNA respectively.  The number of nucleosomes in pulling
experiments by Cui and Bustamante \cite{Cui_pnas_00} is on the
same order ($\approx 300$ nucleosomes).  Each simulation run
consists of 200000 MC sweeps, where one sweep corresponds to
$N_{DNA}$ trials with $N_{DNA}$ being the number of DNA segments.
The amplitudes are chosen such that the acceptance rate equals
approximately 50\%. Every 20 sweeps we save a configuration. As
initial conformation we used the relaxed ($T=0$) fiber structure.
In order to determine the longest relaxation time $\tau_{corr}$ of
the system, we measured the 'time' correlation functions of the
energy, the mass density, the end-to-end distance and the twist.
We typically find that $\tau_{corr}\approx100\mbox{ MC sweeps}$.
Note, however, that strong attractive interactions can cause a
glasslike trapping of the fiber in local energy minima.  As a
consequence, only a limited range of attractive well depths can be
studied meaningfully in simulations. In the present case, the
simulation runs for $\epsilon=4\,k_BT$ were clearly non-ergodic
for low stretching forces. In this case we averaged the results
over 10 independent simulation runs.

The calculation of the elastic constants of the simulated fiber is
performed as follows. First we determine the fiber axis by
calculating the centers of mass $\vec{c}_i = (1/N_c)
\sum_{j=i}^{i+N_c} \vec{R_j}$ of groups of $N_c$ neighboring
nucleosomes. $N_c$ is chosen to match approximately one or two
helical turns. Then we calculate the autocorrelation function
$\langle\mathbf{t}_i\cdot\mathbf{t}_{j}\rangle$ of the tangent
vectors of the fiber axis
\begin{equation}
  \mathbf{t}_i =
  \frac{\vec{c}_{i+1}-\vec{c}_i}{|\vec{c}_{i+1}-\vec{c}_i|}\\
\end{equation}
and extract the persistence length of the fiber $l_p$
from an exponential fit
\begin{equation}
  \langle\mathbf{t}_i\cdot\mathbf{t}_{j}\rangle = \exp
  \left(
    -\frac{|i-j|b}{l_p}
  \right).
\end{equation}
The stretching modulus $k_B T \gamma$ can be estimated via
\begin{equation}
\label{eq:stretchm}
  \gamma = \frac{\langle L\rangle}{\langle\Delta L\rangle^2}
\end{equation}
with $\Delta L=L-\langle L\rangle$ being the mean deviation from
the average contour length of the fiber (defined as the length of
the fiber axis $L=\sum_{i=1}^{N-N_c}|\vec{c}_{i+1}-\vec{c}_i|$).
It should be noted that depending on $N_c$ the estimated values of
the contour length, the persistence length and the stretching
modulus vary. In case of $N_c$ being too large the stretching
modulus is underestimated since bending fluctuations within $i$
and $i+N_c$ are averaged out so that the contour length of the
fiber appears to be smaller. On the other hand, values of $N_c$
which are too small lead to a helicoidal fiber axis, and the
contour length of the fiber is overestimated. This entails a
systematic error which must be minimized. We found that values of
$N_c$ corresponding to one or two helical turns lead to reasonable
estimates.

\section{Results}
\label{sec:results}

\begin{figure}[t]
 \begin{center}
   \includegraphics[angle=0,width=1.0\linewidth]{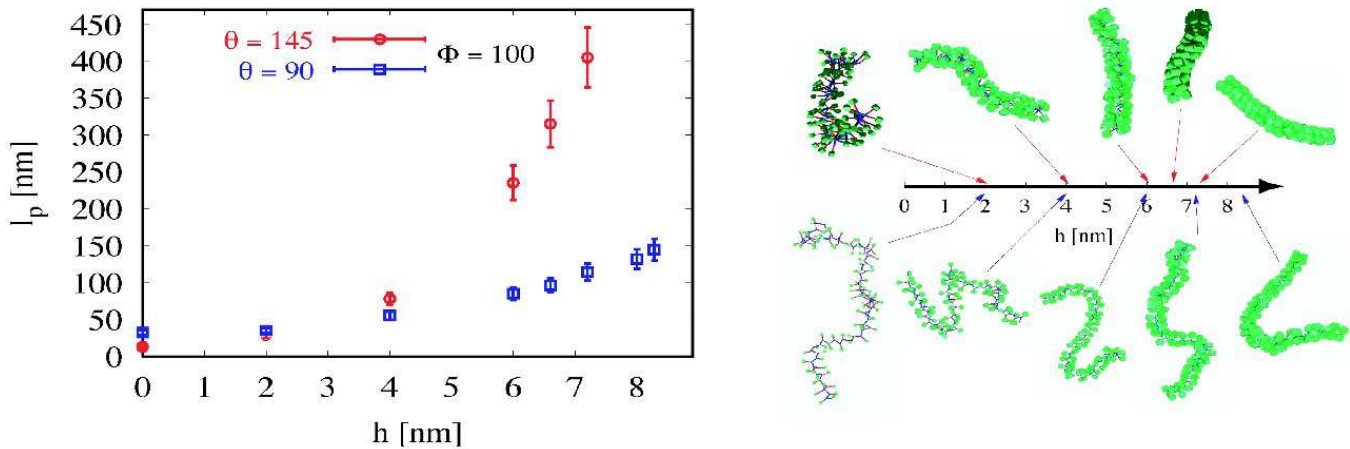}
 \end{center}
 \caption{Effect of the excluded volume interaction on the
   bending persistence length of the fiber for two deflection angles
   $\theta$ with $\phi=100^{\circ}$ and $B=7.14$ nm. For a given
   linker length $B$ the fiber persistence length $l_p$ [$\mbox{nm}$]
   grows with increasing disk size. For very small disk sizes, the measured persistence
   lengths converge to the analytical result, Eq.~(97) of Ref.~\cite{Helmut_03}.
   On the rhs we show some snapshots of simulated fibers with varying disk sizes
   for $\theta=145^{\circ}$ (top row) and $\theta=90^{\circ}$ (bottom row).}
 \label{fig:lpEV}
\end{figure}

In order to quantify the influence of the nucleosome excluded volume on the
fiber stiffness we performed a series of simulations with purely repulsive
interactions between the nucleosome core particles, constant linker geometry
and variable core particle size. The lhs of Fig. \ref{fig:lpEV} shows our
results for the persistence length for two fiber geometries; the corresponding
snapshots in the rhs of the figure illustrate the observed fiber stiffening
with increasing core particle size.

To be more specific, we studied two different sets of angles,
namely $\theta = 145^{\circ}$ and $\phi = 100^{\circ}$ as well as
$\theta = 90^{\circ}$ and $\phi = 100^{\circ}$. In both cases,
starting at $h=0$ (i.e., no nucleosomes present) we observe the
theoretically expected values for the persistence length of the
linker backbone (Eq.~(97) in Ref.~\cite{Helmut_03}), namely $ l_p
\simeq 14$ nm and $ l_p \simeq 34$ nm, respectively. We verified
this formula also for many other values of $\theta$ and $\phi$
(data not shown). However, with increasing $h$ the value of $l_p$
increases and reaches values of $240$ nm and $90$ nm,
respectively, at the canonical value of $h=6$ nm.  This
corresponds to a 17-fold and 3-fold increase respectively of the
fiber stiffness relative to the theoretical prediction. The
different degrees of stiffness (and stiffening) directly reflect
the different nucleosome densities in the two cases: in the
$\theta =145^{\circ}$-fiber the nucleosomes are always quite close
to each other whereas the $\theta=90^{\circ}$-fiber is still much
more open.

\begin{figure}[t]
 \begin{center}
   \includegraphics[angle=0,width=0.5\linewidth]{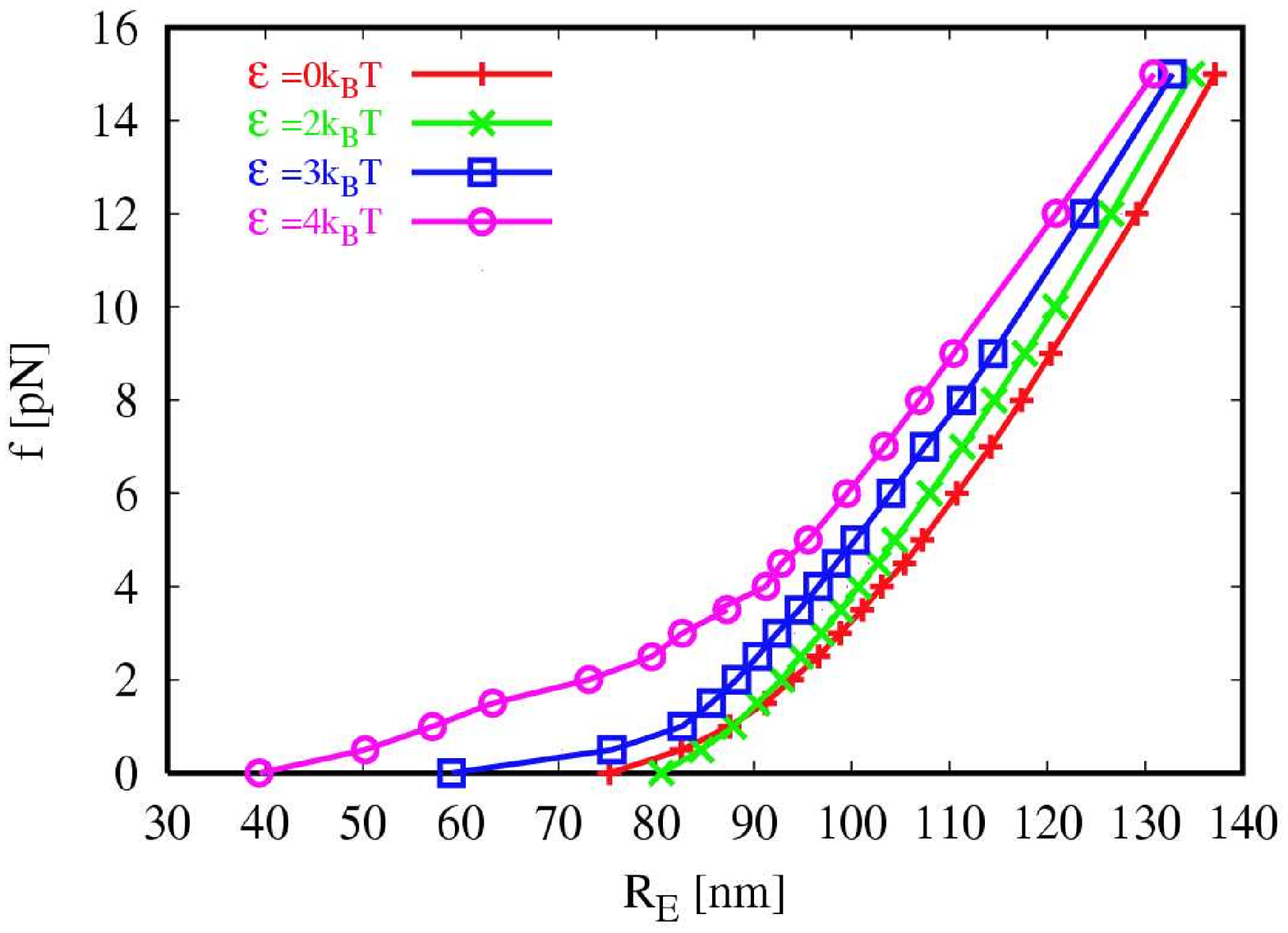}
   \includegraphics[angle=0,width=0.35\linewidth]{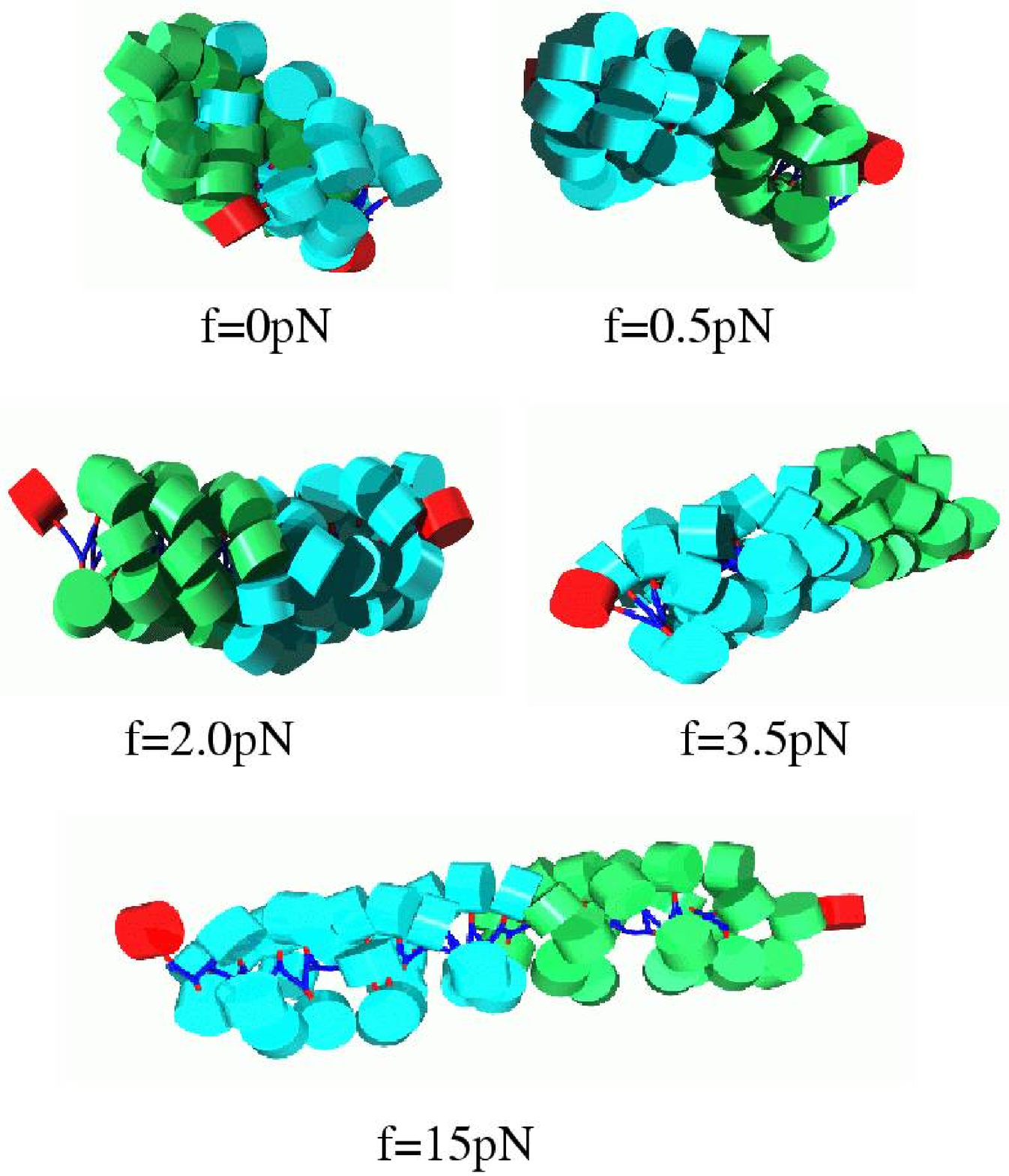}
 \end{center}
 \caption{Force-extension curves for fibers with
   $\theta = 145^{\circ }$, $\phi = 110^{\circ}$ and $B = 7.14$ nm.
   The curves correspond to fibers with different nucleosomal
   attraction, namely $\epsilon=0$ (pure hardcore) and $\epsilon=\,2,\,3,\,4\,k_BT$.
   At $\epsilon=4 k_B T$ occurs a force plateau around $2 pN$.
   Snapshots of fibers with $\theta=145^{\circ}$,
   $\phi=110^{\circ}$, and $B=7.14$ nm at different
   stretching forces for $\epsilon=4\,k_BT$. To facilitate the
   detection of hairpins one half of the chain is
   shown in green while the other half is shown in cyan. The end
   nucleosomes are labelled red. The fiber at $f=0$ pN shows
   a kink close to the center of the chain. Up to $f=2$ pN
   the kink is still present but the ends get pulled out. For $f=15$
   pN the fiber is stretched and nucleosomal contacts are broken.}
 \label{fig:chrompull}
\end{figure}

\begin{figure}[t]
 \begin{center}
   {\includegraphics[angle=0,width=0.39\linewidth]{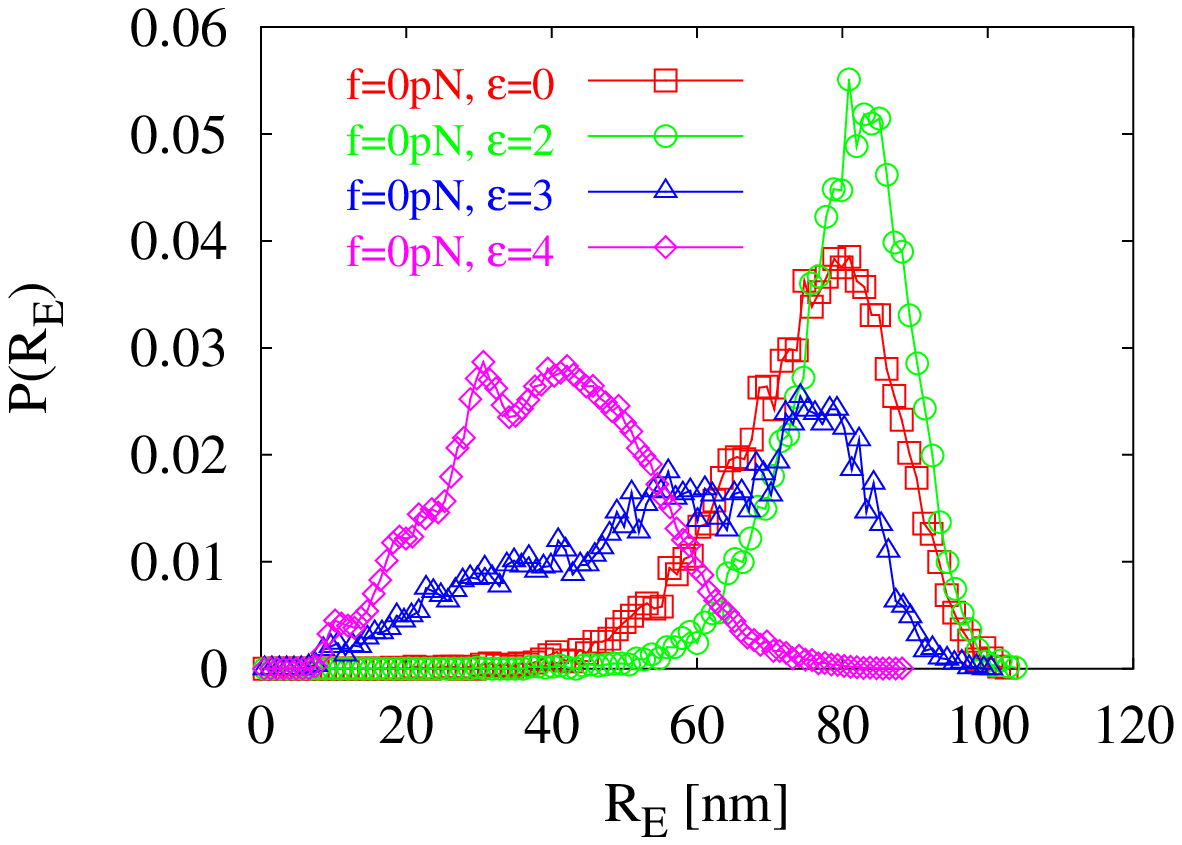}}
   {\includegraphics[angle=0,width=0.39\linewidth]{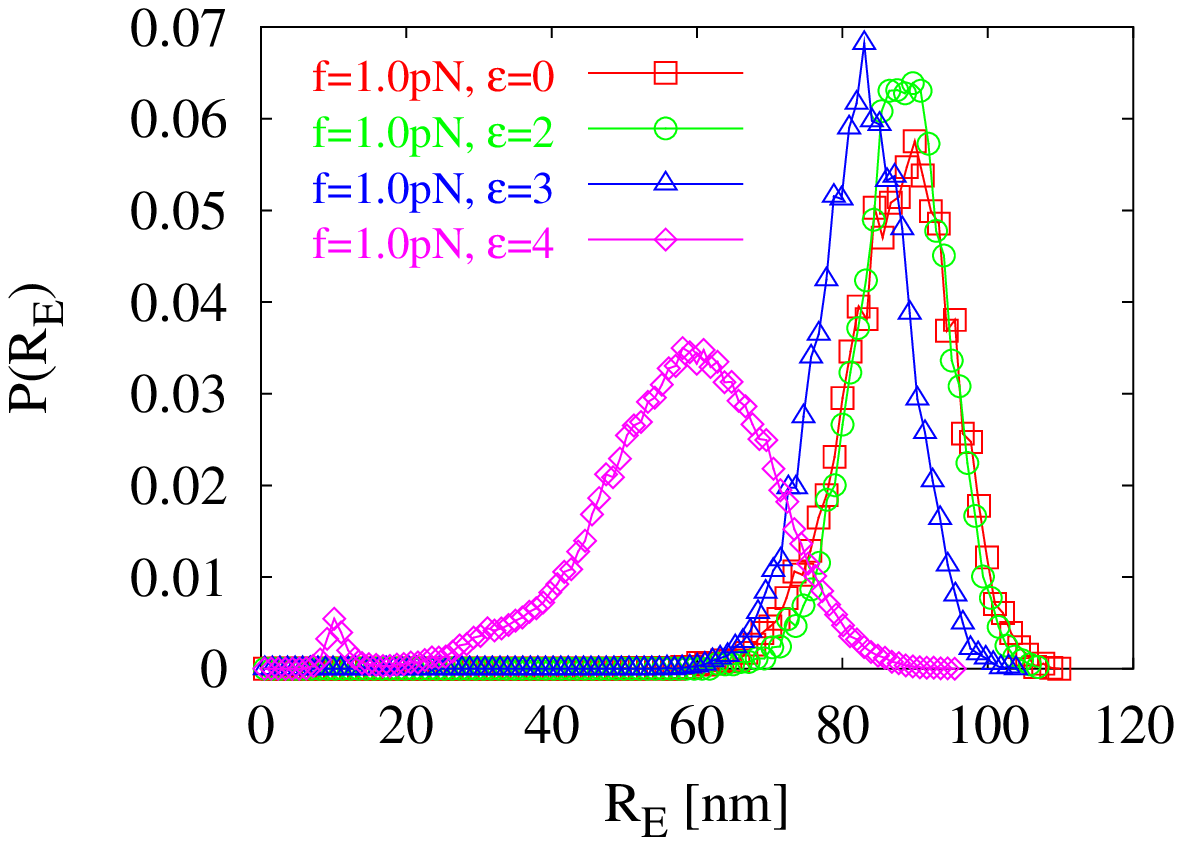}}
   {\includegraphics[angle=0,width=0.39\linewidth]{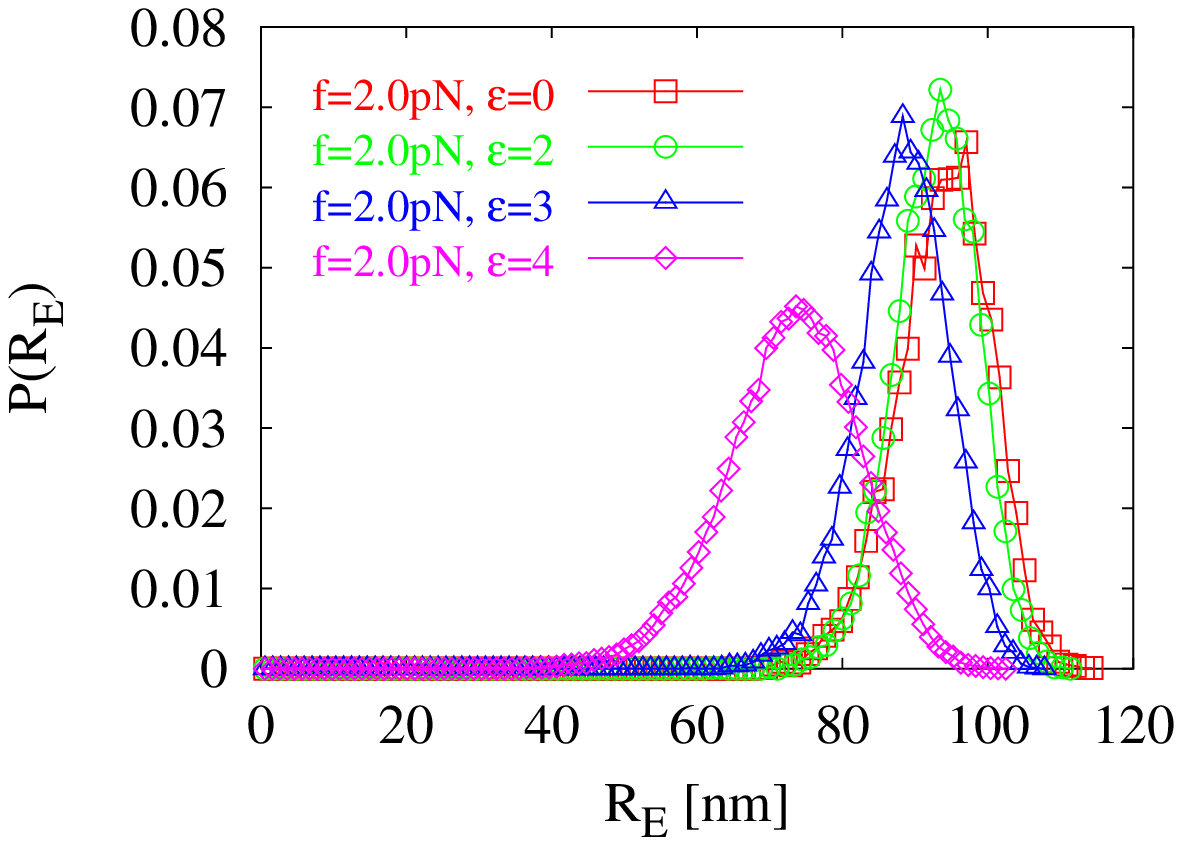}}
   {\includegraphics[angle=0,width=0.39\linewidth]{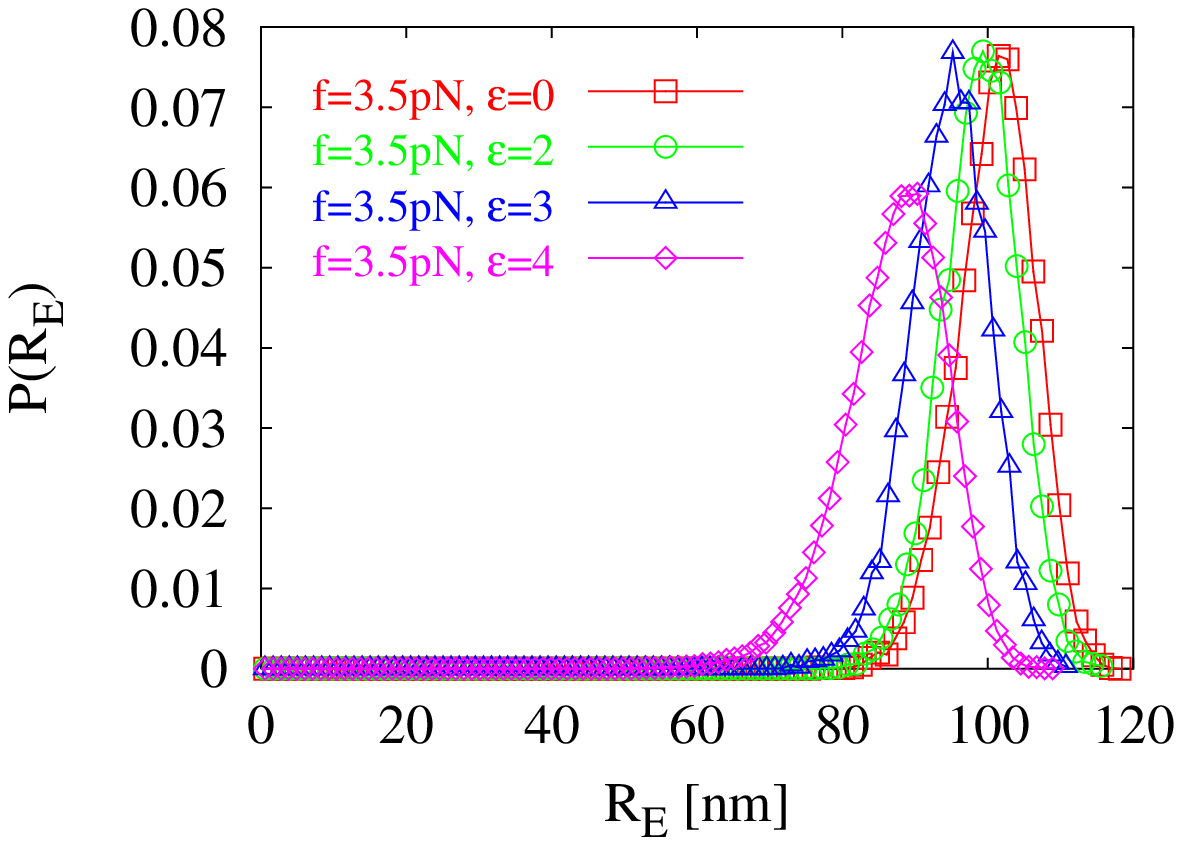}}
    \end{center}
 \caption{Probability density of the end-to-end distance $R_E$ of a fiber
   with $\theta=145^{o}$, $\phi=110^{o}$, and $B=7.14$ nm for various GB
   energy well depths $\epsilon$ and stretching forces $f$ as specified
   in the legends. See text for details.
   }
 \label{fig:repdf}
\end{figure}

We also performed simulations of stretched fibers with $\theta =
145^{\circ }$ and $\phi = 110^{\circ}$.  The resulting
force-elongation curves (red symbols in Fig.~\ref{fig:chrompull})
and histograms of the end-to-end distance distribution (red
symbols in Fig.~\ref{fig:repdf}) show the typical behavior of
extensible worm-like chains. The entropic small force regime is
strongly suppressed, since the $f=0$ contour length, $L=85$nm, of
fibers containing 50 nucleosomes is much smaller than their
persistence length of $l_p=220$ nm. The effective stretching
modulus $\gamma =8$nm$^{-1}$ at finite extensions is smaller than
the value $\gamma =14$nm$^{-1}$ deduced from the analysis of the
length fluctuations (see Eq.~\ref{eq:stretchm}) and larger than
the the theoretical prediction  $\gamma =3$nm$^{-1}$ which is
based on the linker mechanics alone.

For our study of the effect of attractive interactions between
core particles we focused on linker backbone geometries which by
themselves already lead to relatively dense fibers ($\theta =
145^{\circ }$ and $\phi = 110^{\circ}$). As a consequence, the
local fiber geometry remains basically unchanged when we vary the
strength of the attractive interaction from $\epsilon=0$ (the case
of purely repulsive interactions discussed above) to $\epsilon=4$.
For example, the observed attraction induced reductions in the
fiber contour lengths are only of the order of 10\%.

Weak attraction up to $\epsilon=2k_BT$ (green symbols in
Figs.~\ref{fig:chrompull} and \ref{fig:repdf}) has only a small
effect on the observed fiber properties.  While the contour length
decreases to about $L=80$nm, there is a corresponding small
increase in both the fiber stiffness and its stretching modulus,
yielding an overall small increase in the end-to-end distance. The
situation changes dramatically for larger values of the attractive
well depth. In the absence of external forces fibers with
$\epsilon=3k_BT$ and, in particular, with $\epsilon=4k_BT$
(blue/magenta symbols in Figs.~\ref{fig:chrompull} and
\ref{fig:repdf}) have considerably smaller and more broadly
distributed end-to-end distances $R_E$. For small forces the
$\epsilon=4k_BT$ fiber shows a quasi-plateau in the
force-extension curve.

The origin of this peculiar behavior can be identified by
inspecting snapshots of the $\epsilon=4k_BT$ fiber
(Fig.~\ref{fig:chrompull}): At $f=0$ pN the fiber forms a hairpin
structure where the end nucleosomes (in red) are located at the
same end of the structure. The hairpin persists at $f=0.5$ pN but
opens up into a straight configuration around $f=2$ pN. For larger
forces all observed force elongation curves show qualitatively the
same WLC behavior.

\section{Discussion}
\label{sec:discussion}

Our results show that the properties of dense chromatin fibers are
dominated by the interactions between the nucleosome core
particles, while the mechanical properties of the linker backbone
play a less important role. The dominant effect of excluded volume
interactions is to stiffen the fiber. The persistence lengths of
the order of 200 - 250 nm observed in the present study are in
good agreement with the values obtained by Langowski et al.
\cite{Munkel_pre_98,Wedemann_bpj_02}. The surprising result of our
simulations is that attractive interactions of a few $k_BT$ per
nucleosome pair are sufficient to bend fibers with a contour
length of about $1/3$ of the persistence length into dense hairpin
configurations. Judging from Fig.~5 in Ref.~\cite{Katritch_jmb_00}
hairpins also occurred in the simulations by Katritch et al. --
even though the authors did not discuss this issue in their paper.
Since the observed force-elongation curves show qualitatively
similar features in all three cases, it is tempting to speculate
that similar effects also occurred in the stretching experiment by
Cui and Bustamante \cite{Cui_pnas_00}. In the following, we will
discuss the condensation of semi-flexible filaments,
the local structure of the hairpins observed in our
simulations, the signature of hairpin opening in stretching
experiments, and possible implications of our observations for the
folding of chromatin into chromosomes.

\subsection{Condensation of semi-flexible filaments}

The condensation of semi-flexible filaments due to short-range
attractive interactions has recently been treated in detail by
Schnurr et al. \cite{Schnurr_pre_02}. They introduced a {\em
condensation length} $L_c=\sqrt{k_BT l_p/\sigma_{attr}}$
($\sigma_{attr}$: attraction energy per length) by balancing
expressions for typical bending and surface energies, $k_BT
l_p/L\sim \sigma_{attr}L$. Filaments with a contour length $L/L_c<
{\cal O}(10)$ remain extended, while longer chains aggregate into
structures with a typical size of the order of $L_c$. So-called
``racquet'' states consist of a straight stem where the chain is
folded back on itself several times and where $180^\circ$ U-turns
are shaped like the head of a tennis racquet. A racquet state with
one turning point resembles our hairpin structure. Longer fibers
lower their energy by having multiple turning points. For
infinitely thin filaments Schnurr et al. were able to show that
the kinetically preferred racquet states \cite{Schnurr_epl_00}
have slightly higher energies than toroidal structures.

In the present case, we measured a typical attraction energy per
length of $\sigma_{attr} \approx \left( 3/4 \right) k_{B} T
nm^{-1}$ for hairpin conformations of the $\epsilon=4 k_BT$-fiber
in good agreement with an estimate of $\epsilon/h = 4 k_{B} T/6
nm$ for the attractive energy per nucleosome pair of height $h$.
Using this estimate and a persistence length $l_p=240$nm,
chromatin fibers with weak attractive interactions should start to
condense, if they are longer than about ${\cal O}(10)L_c \approx
\sqrt{k_BT/\epsilon}$ 400 nm. Note that the above argument breaks
down for $\epsilon > 6k_BT$, when the predicted racquet head or
torus radii become smaller than the radius of the 30 nm-fiber.

Can the structure of the hairpins we observe really be understood
in the framework of the WLC model?  Or does one have to consider
the relatively sharp turning points with the curvature radii of
the order of the fiber radius as localized kink defects? Following
the analysis of Schnurr et al. \cite{Schnurr_pre_02}, the fibers
in our simulations are too short to fold back on themselves. The
WLC estimate of the bending energy for a U-turn with radius 15nm
is $U_{kink}\approx25k_BT$, while the onset of hairpin formation
at $\epsilon=3 k_BT$ in 80 nm long fibers suggest that the actual
energy penalty for the kinks is as small as
$U_{kink}\approx9k_BT$.  We note that the combination of finite
fiber radii and relatively small kink energies should stabilize
racquet structures relative to toroids.

\subsection{Hairpin structure}

\begin{figure}[t]
 \begin{center}
   {\includegraphics[angle=0,width=0.33\linewidth]{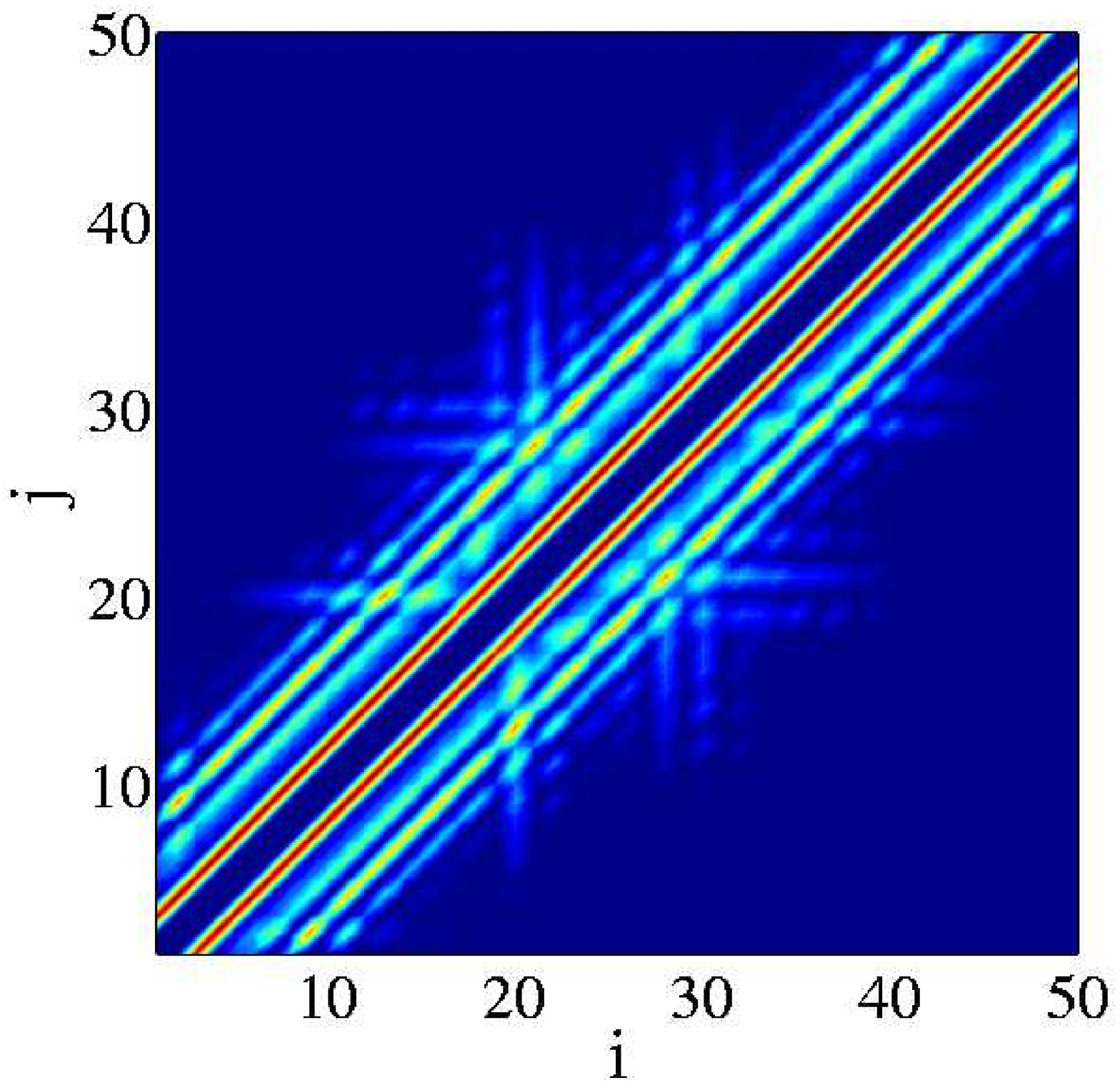}}
   {\includegraphics[angle=0,width=0.33\linewidth]{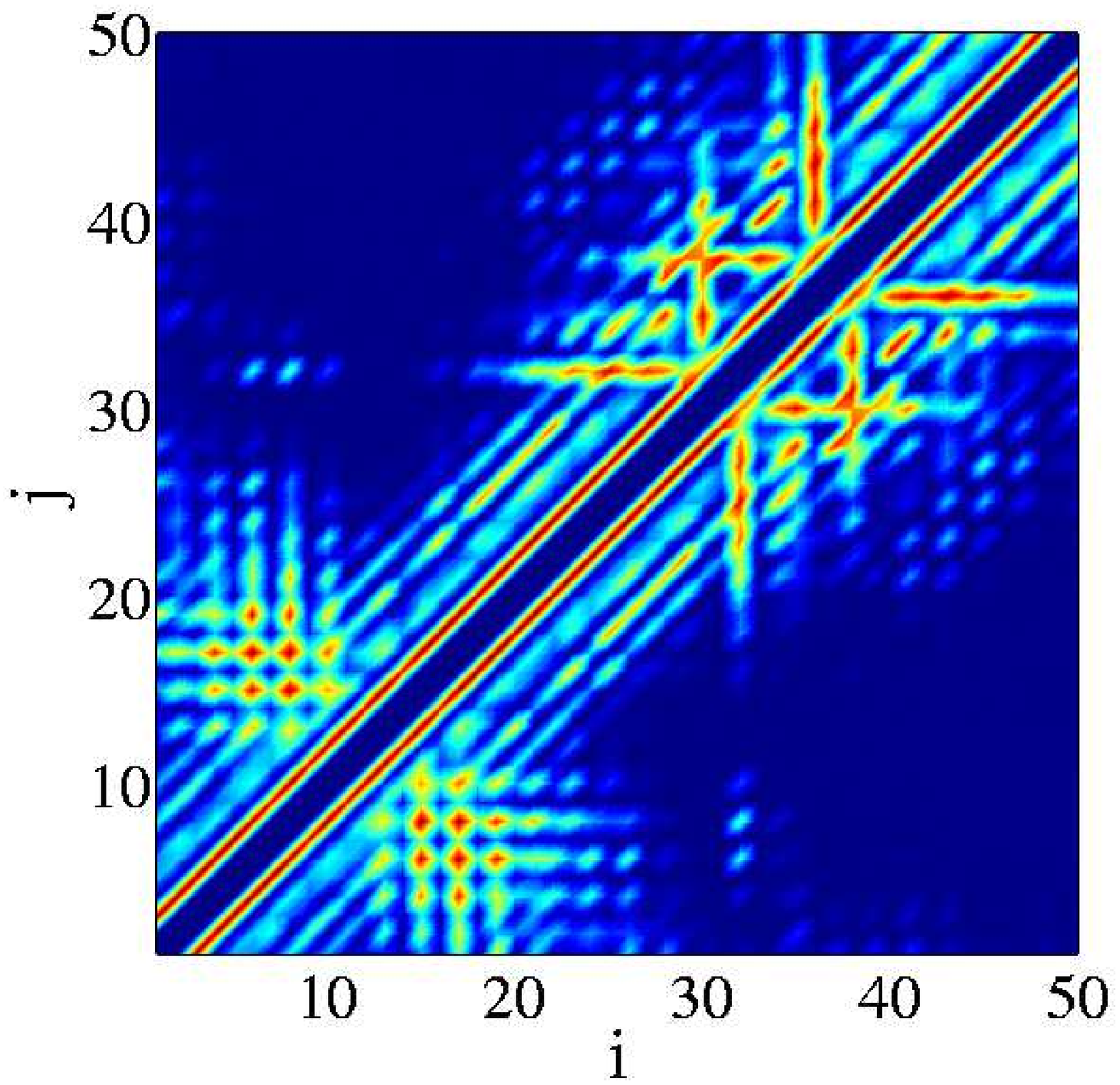}}
   {\includegraphics[angle=0,width=0.33\linewidth]{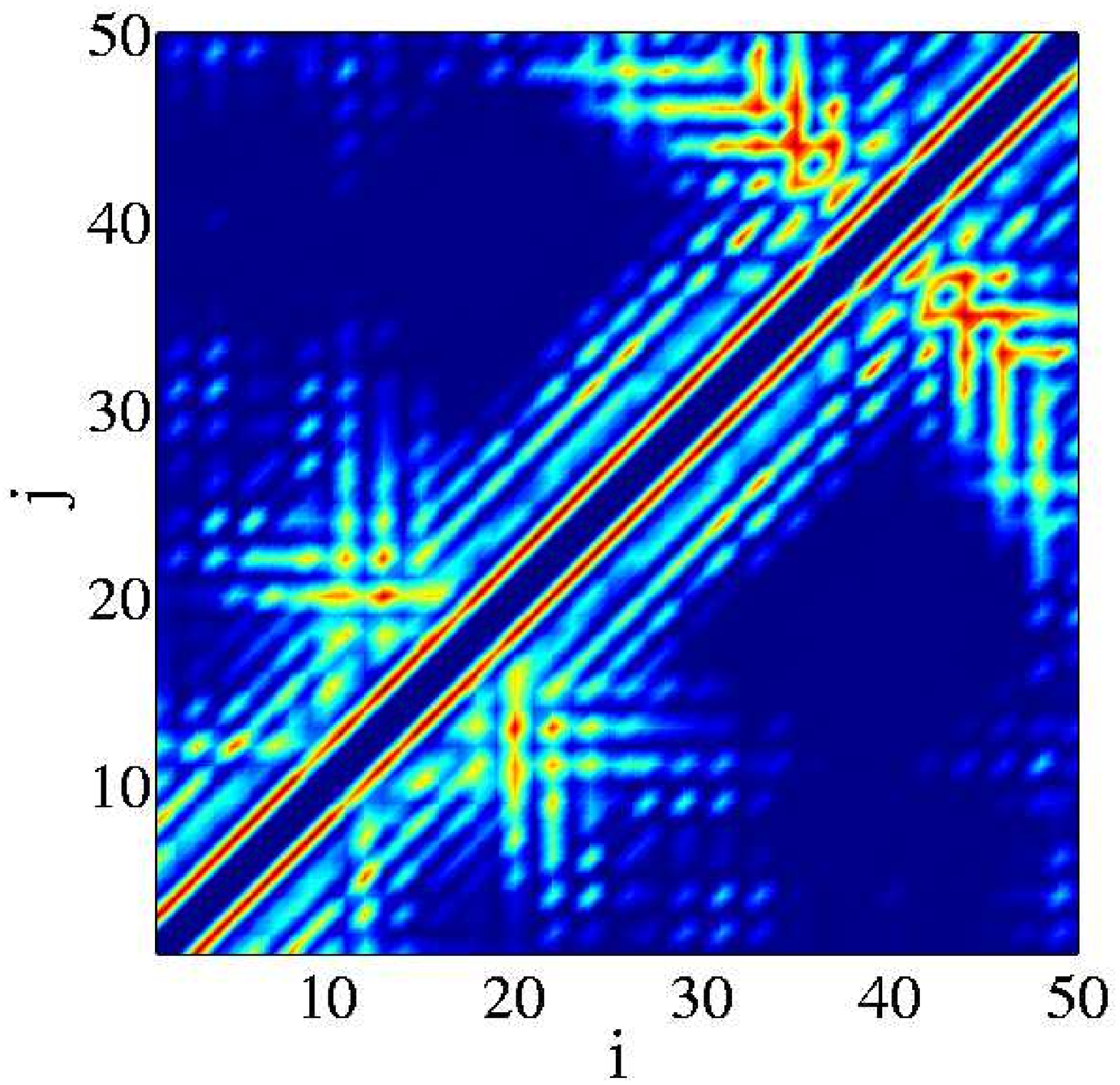}}
   {\includegraphics[angle=0,width=0.33\linewidth]{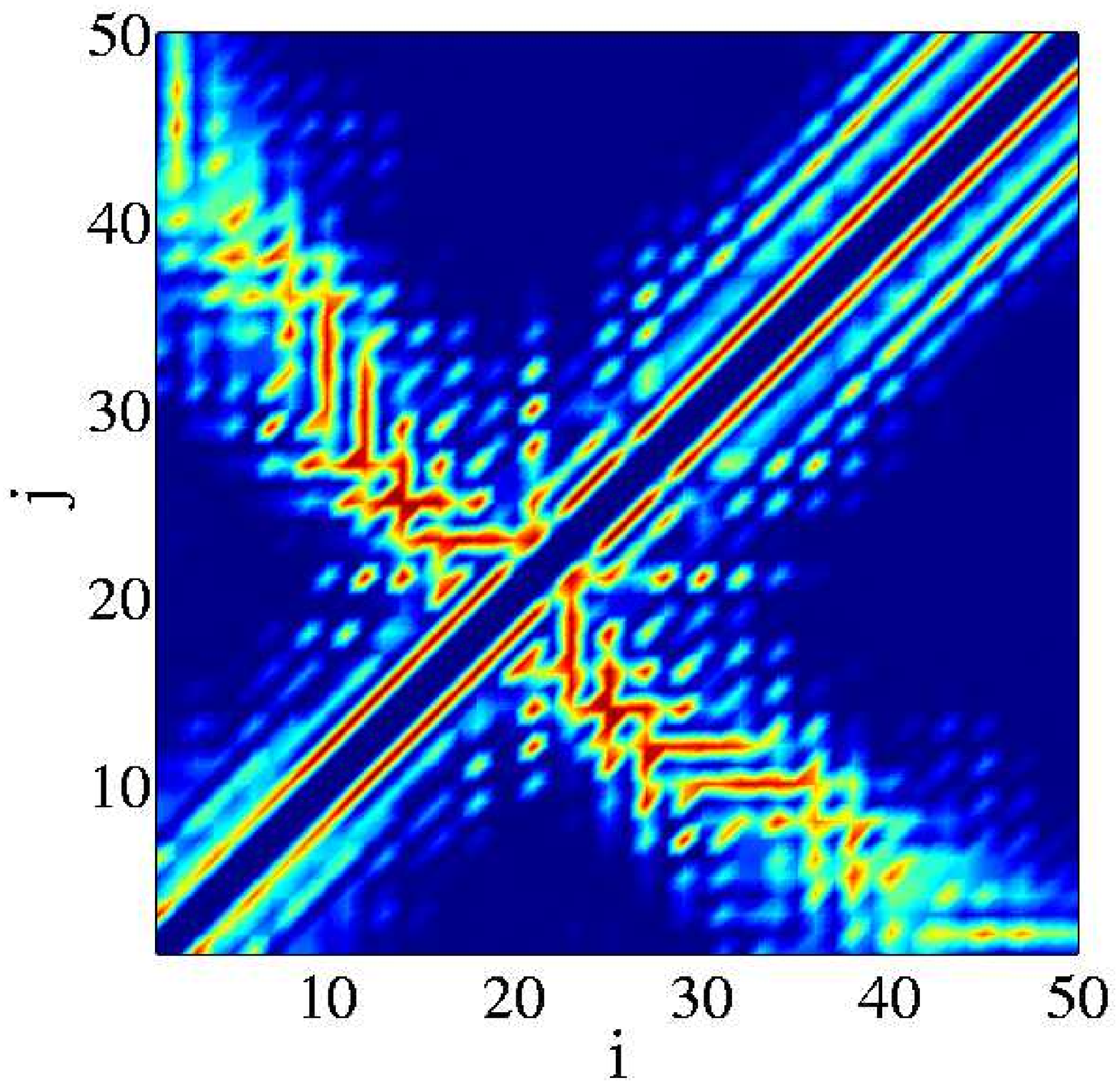}}
 \end{center}
 \caption{Contact matrix of the fiber with $\epsilon=4\,k_BT$,
   $\theta=145^{\circ}$, $\phi=110^{\circ}$, and $B=7.14$ nm.
   (a) At an external tension $f=3.5$ pN the fiber is stretched
   and there is no hairpin. (b) to (d) Without an external force,
   $f=0$ pN, one can clearly detect hairpin structures. The simulation
   runs (b) and (c) show the occurrence of two hairpins whereas
   in (d) only one hairpin can be identified.}
 \label{fig:contmat}
\end{figure}

The internal structure of hairpin configurations shows up very
clearly in the contact matrix ${\cal M}_{contact}$ of the fiber
that we depict in Fig.~\ref{fig:contmat}. For a given
configuration the contact matrix is defined as follows: If a pair
of nucleosomes $i$ and $j$ is in contact (center-to-center
distance smaller than $1.5 d= 15nm$) one has ${\cal
M}^{(ij)}_{contact}=1$, otherwise ${\cal M}^{(ij)}_{contact}=0$.
By taking the average of the matrix elements for many
configurations of a given simulation run we obtain two-dimensional
histograms as it is shown in Fig.~\ref{fig:contmat}. For large
forces no kinks are present and the nucleosomes of the fiber form
a quite regular crossed-linker structure. We find two pronounced
diagonals at the positions $i=j \pm 2$ that correspond to
"short-ranged" excluded volume interaction. Less pronounced side
stripes parallel to the main diagonal are also observed and
describe interactions between nucleosomes of neighboring turns,
especially at $i=j \pm 5$ and at $i=j \pm 7$. Hairpins manifest
themselves by crosslike patterns in the contact matrices with
branches along the secondary diagonal. The location of the kink
corresponds to the point where the two diagonals cross each other.
Closer inspection of the histograms shows that the hairpin arms
outside the kink region are partially interdigitated but not
strongly perturbed. However the effect is strong enough to
effectively prevent any possibility of sliding of the two arms
with respect to each other on the time scale of our simulations.
Therefore a hairpin structure, once formed, can be very stable
with a frozen in position of the turning point. With respect to
the fiber structure at the turning point,
Fig.~\ref{fig:contmat}(d) in particular shows disruptions of the
internal order of the nucleosomes: The diagonals $i=j \pm 2$ are
broken at that position, presumably due to broken contacts of
nucleosomes located at the outside of the turning point. This
observation supports the picture of a localized kink defect.

\subsection{Large scale vs. local decondensation
in stretching experiments}

Toroidal or racquet-like condensates become unstable under the
influence of a stretching force, if the applied force $f$ exceeds
a value on the order of the attractive energy per unit length,
$\sigma_{attr}$. Neglecting kinetic barriers (for a discussion of
DNA unwrapping from nucleosomes or toroidal aggregates see
\cite{kulic_04}), the corresponding force plateau in the
force-extension curves for chromatin fibers should be observed for
forces of $f\approx\sigma_{attr}\approx (2/3) (\epsilon/k_BT)$pN.
In order to experimentally distinguish this scenario from local
structural changes which are expected to occur at similar force
levels \cite{Schiessel_bpj_01}, one should plot force-{\em
relative} extension curves for fibers reconstituted under similar
conditions on DNA strands of different length. In the case of
local structural changes, the obtained curves coincide, while for
a global decondensation the size of the unperturbed state is of
order $L_c$ independently of the fiber length.

\subsection{Biological implications}

As a final point we discuss possible biological implications of
fiber stiffening and hairpin formation. Both effects are
associated with characteristic length scales; fiber stiffening
with a persistence length on the order of 300 nm, hairpin
formation with a condensation length on the order of a few tens of
nm.

We note again that the biochemical control of the effective
interaction between nucleosomes is essential for the condensation
and decondensation of chromosomes during the cell cycle
\cite{Horn_sci_02}. There is indirect experimental evidence for
the formation of chromosomal DNA loops having a size of about 50
kilobases \cite{CalladineDrew99} (corresponding to about 250
nucleosomes), e.g. from the comparison of separation patterns of
excised large DNA fragments by pulsed field electrophoresis with
the patterns obtained by DNA cleavage by topoisomerase II of the
nuclear matrix of the cell \cite{Razin_jbc_95,Razin_pnas_95}.
Using a nucleosome density of 6/(11 nm), 250 nucleosomes
correspond to a fiber length of 460 nm, i.e. about twice the
persistence length of 240 nm found in our simulations.

There are two scenarios linking the persistence length to the loop
size; both are based on the observation that the free energy cost
for the formation of a loop is smallest for semi-flexible chains
with a length of about twice the persistence length. For shorter
chains the bending energy is high, for larger chains the entropy
of chain conformations counteracts loops formation. In the first
(equilibrium) scenario, the chromatin fiber loops around strongly
attracting organizing centers. Spherical centers could induce the
formation of rosette structures \cite{Schiessel_epl_00}.
Filamentous organizing centers could lead to structures resembling
cartoons of chromosomes often found in biological textbooks, cf.
e.g. Ref.~\cite{cell}. The second scenario which links the
persistence length to the loop size is closer to what we observed
in our simulations: if the attraction between fibers is strong
enough to prevent structural reorganization after the two halves
of the fiber have tightly closed the gap between the point of
first contact and the hairpin defect, then loops with the typical
size of the fiber persistence length are preferred for kinetic
reasons. Hairpins have indeed been observed in cryo-EM pictures of
chromatin fibers in the presence of MENT \cite{Grigoryev_jbc_99},
a protein that is involved in the formation of dense,
transcriptionally inert sections of chromatin, the so-called
heterochromatin.

If, on the other hand, structural reorganization is still possible
in collapsed fibers, then the size of the aggregates should be
controlled by the smaller condensation length and not by the
persistence length of the 30-nm fiber. Fibers with diameters
ranging between 60 to 130 nm diameter have indeed been observed in
mitotic and in G1 chromosomes and are called chromonema fibers
\cite{Belmont_jcb_94}, cf. also Ref.~\cite{Horn_sci_02}. However,
the folding of 30-nm fibers into superfibers with a diameter on
the order of the condensation length requires a mechanism that
prevents the formation of globular aggregates.

\section{Conclusion}
\label{sec:conclusion}

We have used computer simulations of model chromatin fibers to investigate the
influence of excluded volume and attractive interactions between the
nucleosomes on the large-scale structure and elasticity of the 30 nm chromatin
fiber. Our results shed light ({\it i}) on the discrepancy between the
theoretically expected and the observed persistence length of 200-300 nm as
well as ({\it ii}) on the somewhat counterintuitive observation that such
fibers nevertheless seem to be able to curl up into chromonema fibers with
diameters of 60 - 130 nm.

Our results clearly show that the stiffness of dense fibers is dominated by
excluded volume interactions between nucleosomes. The observed persistence
lengths exceed estimates based on the linker backbone elasticity by one order
of magnitude. With respect to internucleosomal attraction we have concentrated
on generic, non-local aspects of fiber condensation. The 2 pN force plateau
observed in our simulations reflects a structural feature {\it
beyond} the 30 nm fiber: the opening of a hairpin. Our results suggest that
aligned (anti-) parallel fibers are only weakly perturbed, but that the fiber
geometry is locally disrupted at the $180^{\circ}$ turning point. The crucial
point is that the energetic cost of this localized defect is considerably
smaller than for a smoothly bending semi-flexible filament with the
same bending persistence length as chromatin. This effect could be important
for an understanding of chromatin folding in chromosomes.

\section{Acknowledgments}

We thank K. Kremer, I. M. Kulic, J. Langowski and J. Widom for
valuable discussions, M. R. Ejtehadi for providing parts of the
simulation code and a referee for useful comments on how to
structure the material presented in a first draft of the
present manuscript. BM and RE gratefully acknowledge financial
support from an Emmy-Noether grant of the DFG.


\end{document}